\documentclass[conference]{IEEEtran}
\IEEEoverridecommandlockouts

\usepackage{cite}
\usepackage{amsmath,amssymb,amsfonts}
\usepackage{algorithmic}
\usepackage{graphicx}
\usepackage{textcomp}
\usepackage{xcolor}
\def\BibTeX{{\rm B\kern-.05em{\sc i\kern-.025em b}\kern-.08em
    T\kern-.1667em\lower.7ex\hbox{E}\kern-.125emX}}

\usepackage{booktabs} 
\usepackage[nolist,nohyperlinks]{acronym}

\usepackage{makecell}

\usepackage{array}

\usepackage{xurl}

\acrodef{ICS}[ICS]{Industrial Control System}
\acrodefplural{ICSs}[ICS]{Industrial Control Systems}
\acrodef{NMA}[NMA]{Network Management Agent}
\acrodefplural{NMAs}[NMA]{Network Magement Agent}
\acrodef{RQ}[RQ]{Research Question}
\acrodefplural{RQs}[RQ]{Research Questions}
\acrodef{ML}[ML]{Machine Learning}
\acrodef{PLC}[PLC]{Programmable Logic Controller}
\acrodefplural{PLCs}[PLC]{Programmable Logic Controllers}
\acrodef{HMI}[HMI]{Human Machine Interface}
\acrodefplural{HMIs}[HMI]{Human Machine Interfaces}
\acrodef{PHY}[PHY]{Physical Layer device}
\acrodefplural{PHYs}[PHY]{Physical Layer devices}
\acrodef{API}[API]{Application Programming Interface}
\acrodefplural{APIs}[API]{Application Programming Interfaces}
\acrodef{ICN}[ICN]{Industrial Control Network}
\acrodefplural{ICNs}[ICN]{Industrial Control Networks}
\acrodef{DPI}[DPI]{Deep Packet Inspection}
\acrodef{MTBF}[MTBF]{Mean Time Between Failures}
\acrodef{IEEE}[IEEE]{Institute of Electrical and Electronics Engineers}
\acrodef{IETF}[IETF]{Internet Engineering Task Force}
\acrodef{RFC}[RFC]{Request for Comment}
\acrodefplural{RFCs}[RFC]{Request for Comments}
\acrodef{NIC}[NIC]{Network Interface Card}
\acrodefplural{NICs}[NIC]{Network Interface Cards}
\acrodef{OS}[OS]{Operating System}
\acrodefplural{OSs}[OS]{Operating System}
\acrodef{IOT}[IoT]{Internet of Things}
\acrodef{PROFIBUS}[PROFIBUS]{Process Field Bus}
\acrodef{DP}[DP]{Decentralised Peripherals}
\acrodef{CAN}[CAN]{Controller Area Network }
\acrodef{PROFINET}[PROFINET]{Process Field Network}
\acrodef{OPC}[OPC]{Open Platform Communications}
\acrodef{UA}[UA]{Unified Architecture}
\acrodef{EtherNet}[EtherNet/IP]{EtherNet/Industrial Protocol}
\acrodef{IEC}[IEC]{International Electrotechnical Commission}
\acrodefplural{RPC}[RPCs]{Remote Procedure Calls}
\acrodef{RTU}[RTU]{Remote Terminal Unit}
\acrodef{SDN}[SDN]{Software-Defined Networking}
\acrodef{SSH}[SSH]{Secure Shell Protocol}
\acrodef{FTP}[FTP]{File Transfer Protocol}
\acrodef{HTTP}[HTTP]{Hypertext Transfer Protocol}
\acrodef{CLI}[CLI]{Command-Line Interface}
\acrodef{FPGA}[FPGA]{Field-Programmable Gate Array}
\acrodef{CDP}[CDP]{Cisco Discovery Protocol}
\acrodef{DTP}[DTP]{Dynamic Trunking Protocol}
\acrodef{MDNS}[mDNS]{Multicast Domain Name System} 
\acrodef{PID}[PID]{Process identifier}
\acrodef{SCP}[SCP]{Secure Copy Protocol}
\acrodef{USB}[USB]{Universal Serial Bus}
\acrodef{JSON}[JSON]{JavaScript Object Notation}
\acrodef{RT}[RT]{real-time}
\acrodef{TSN}[TSN]{Time-Sensitive Networking}
\acrodef{CSMA/CD}[CSMA/CD]{Carrier Sense Multiple Access/Collision Detection}
\acrodef{PTP}[PTP]{Precision Time Protocol}
\acrodef{NTP}[NTP]{Network Time Protocol}
\acrodef{CNC}[CNC]{Centralized Network Configuration}
\acrodef{SRP}[SRP]{Stream Reservation Protocol}
\acrodef{MRP}[MRP]{Media Redundancy Protocol}
\acrodefplural{MRP}[MRPs]{Media Redundancy Protocols}
\acrodef{MRM}[MRM]{Media Redundancy Manager}
\acrodefplural{MRM}[MRMs]{Media Redundancy Managers}
\acrodef{MRC}[MRC]{Media Redundancy Client}
\acrodef{STP}[STP]{Spanning Tree Protocol}
\acrodef{RSTP}[RSTP]{Rapid Spanning Tree Protocol}
\acrodef{PRP}[PRP]{Parallel Redundancy Protocol}
\acrodef{BFD}[BFD]{Bidirectional Forwarding Detection}
\acrodef{SNMP}[SNMP]{Simple Network Management Protocol}
\acrodef{MIB}[MIB]{Management Information Base}
\acrodefplural{MIBs}[MIB]{Management Information Bases}
\acrodef{OID}[OID]{Object Identifier}
\acrodefplural{OIDs}[OID]{Object Identifiers}
\acrodef{NMS}[NMS]{Network Management Station}
\acrodef{ISO}[iso]{International Organization for Standardization} 
\acrodef{LLDP}[LLDP]{Link Layer Discovery Protocol}
\acrodef{MED}[MED]{Media Endpoint Discovery}
\acrodef{ICMP}[ICMP]{Internet Control Message Protocol}
\acrodef{ARP}[ARP]{Address Resolution Protocol}
\acrodef{NETCONF}[NETCONF]{Network Configuration Protocol}
\acrodef{TCP/IP}[TCP/IP]{Transmission Control Protocol/Internet Protocol}
\acrodef{TCP}[TCP]{Transmission Control Protocol}
\acrodef{UDP}[UDP]{User Datagram Protocol}
\acrodef{OSI}[OSI]{Open Systems Interconnection}
\acrodef{MAC}[MAC]{Medium Access Control}
\acrodef{IP}[IP]{Internet Protocol}
\acrodef{LAN}[LAN]{Local Area Network}
\acrodef{WAN}[WAN]{Wide Area Network}
\acrodef{MAN}[MAN]{Metropolitan Area Network}
\acrodefplural{LANs}[LAN]{Local Area Networks}
\acrodefplural{WANs}[WAN]{Wide Area Networks}
\acrodefplural{MANs}[MAN]{Metropolitan Area Networks}
\acrodef{SCADA}[SCADA]{Supervisory Control and Data Acquisition} 
\acrodef{VLAN}[VLAN]{Virtual Local Area Network}
\acrodef{VLANs}[VLAN]{Virtual Local Area Networks}
\acrodef{QOS}[QoS]{Quality of Service}
\acrodef{COS}[CoS]{Class of Service}
\acrodef{PCP}[PCP]{Priority Code Point}
\acrodef{DSCP}[DSCP]{Differentiated Services Code Point}
\acrodef{DiffServ}[DiffServ]{Differentiated Services}
\acrodef{IntServ}[IntServ]{Integrated Services}
\acrodef{RSVP}[RSVP]{Resource Reservation Protocol}
\acrodef{RSVP}[RSVP]{Resource Reservation Protocol}
\acrodef{FIFO}[FIFO]{First-In-First-Out}
\acrodef{RR}[RR]{Round-robin}
\acrodef{EF}[EF]{Expedited Forwarding}
\acrodef{OT}[OT]{Operational Technology}
\acrodef{IT}[IT]{Information Technology}
\acrodef{CRC}[CRC]{Cyclic Redundancy Check}
\acrodef{I/O}[I/O]{Input/Output}
\acrodef{CIP}[CIP]{Common Industrial Protocol}
\acrodef{SPOF}[SPoF]{Single Point of Failure}
\acrodefplural{SPOF}[SPoFs]{Single Point of Failures}
\acrodef{DSR}[DSR]{Design Science Research}
\acrodef{DSRM}[DSRM]{Design Science Research Methodology}
\acrodef{BPDU}[BPDU]{Bridge Protocol Data Unit}
\acrodefplural{BPDU}[BPDUs]{Bridge Protocol Data Units}
\acrodef{PVST}[PVST]{VLAN spanning tree protocol}
\acrodef{HSR}[HSR]{High-availability Seamless Redundancy}
\acrodef{RCT}[RCT]{Redundancy Control Trailer}
\acrodef{BER}[BER]{Bit Error Ratio}
\acrodef{DCS}[DCS]{Distributed Control System}
\acrodefplural{DCS}[DCSs]{Distributed Control Systems}
\acrodef{DCN}[DCN]{Distributed Control Node}
\acrodefplural{DCN}[DCNs]{Distributed Control Nodes}
\acrodef{gcc}[gcc]{GNU Compiler Collection}
\acrodef{MIM}[MIM]{Media Redundancy Interconnection Manager}
\acrodef{MIC}[MIC]{Media Redundancy Interconnection Client}
\acrodef{YANG}[YANG]{Yet Another Next Generation}
\acrodef{ACL}[ACL]{Access Control List}
\acrodefplural{ACL}[ACLs]{Access Control Lists}

\acrodef{NHB}[NHB]{node heartbeat}
\acrodefplural{NHB}[NHB]{node heartbeats}

\acrodef{NWHB}[NWHB]{network heartbeat}
\acrodefplural{NWHB}[NWHB]{network heartbeats}

\acrodef{IIC}[IIC]{Industry IoT Consortium}

\acrodef{OPC UA}[OPC UA]{Open Platform Communications Unified Architecture}

\acrodef{ST}[ST]{Scheduled Traffic}
\acrodefplural{ST}[STs]{Scheduled Traffics}

\acrodef{AVB}[AVB]{Audio-Video Bridging}
\acrodefplural{AVB}[AVBs]{Audio-Video Bridgings}

\acrodef{BE}[BE]{Best Effort }
\acrodefplural{BE}[BEs]{Best Effort }

\acrodef{HRT}[HRT]{Hard Real-time}
\acrodef{D}[D]{Deadline}
\acrodef{JI}[JI]{Jitter Input}
\acrodef{JO}[JO]{Jitter Output}
\acrodef{P}[P]{Periodicity}

\acrodef{QCAP}[QCAP]{Quality Checks After Production}

\acrodef{WCRTA}[WCRTA]{Worst-case Response Time Analysis}
\acrodef{WCRT}[WCRT]{Worst-case Response Time}

\acrodef{CBS}[CBS]{Credit-based Shaper}
\acrodef{TAS}[TAS]{Time Aware Shaper}
\acrodef{GCL}[GCL]{Gate Control List}
\acrodef{NG}[NG]{Not Guaranteed}

\acrodef{UA}[UA]{Unified Architecture}
\acrodef{UADP}[UADP]{UA Datagram Protocol}

\acrodef{PubSub}[PubSub]{Publish-Subscribe}

\begin{document}

\title{Traffic-Aware Configuration of OPC UA PubSub in Industrial Automation Networks}

\author{Kasra Ekrad\textsuperscript{1,2}, 
        Bjarne Johansson\textsuperscript{2},
        Inés Alvarez Vadillo\textsuperscript{2}, 
        Saad Mubeen\textsuperscript{1},  
        Mohammad Ashjaei\textsuperscript{1} \\
        \textsuperscript{1}M{\"{a}}lardalen University, V{\"{a}}ster{\aa}s, Sweden\\ \textsuperscript{2}{ABB}, V{\"{a}}ster{\aa}s, Sweden\\

        \{kasra.ekrad, saad.mubeen, mohammad.ashjaei\}@mdu.se,\\
        \{ines.alvarez-vadillo, bjarne.johansson\}@se.abb.com
}

\maketitle



\begin{abstract}
\label{sec:abstract}
Interoperability across industrial automation systems is a cornerstone of Industry 4.0. To address this need, the OPC Unified Architecture (OPC UA) Publish-Subscribe (PubSub) model offers a promising mechanism for enabling efficient communication among heterogeneous devices. PubSub facilitates resource sharing and communication configuration between devices, but it lacks clear guidelines for mapping diverse industrial traffic types to appropriate PubSub configurations. This gap can lead to misconfigurations that degrade network performance and compromise real-time requirements. This paper proposes a set of guidelines for mapping industrial traffic types, based on their timing and quality-of-service specifications, to OPC UA PubSub configurations. The goal is to ensure predictable communication and support real-time performance in industrial networks. The proposed guidelines are evaluated through an industrial use case that demonstrates the impact of incorrect configuration on latency and throughput. The results underline the importance of traffic-aware PubSub configuration for achieving interoperability in Industry 4.0 systems.

\end{abstract}



\section{Introduction}
\label{sec:introduction}

Industrial automation systems are increasingly moving towards converged communication technologies, requiring networks that support a wide range of traffic types with varying criticalities. The OPC Unified Architecture (OPC UA), particularly its Publish-Subscribe (PubSub) extension defined in Part 14 of the specification~\cite{opc_pubsub_2024}, has emerged as a key enabler of scalable and interoperable communication across heterogeneous systems. OPC UA is an open-source service-oriented architecture that enables secure, platform-independent data exchange across heterogeneous devices, from the field to the cloud~\cite{szabo_towards_2024}. Unlike the Client-Server model, which lacks support for real-time communications, the PubSub model provides a foundation for a scalable communication pattern~\cite{lu_proposal_2025}. It promises effective data delivery to multiple recipients by publishing the messages to the message-oriented middleware without publishers and subscribers knowing each other.

Despite the above-mentioned advantages of OPC UA, there is a lack of systematic guidance for configuring OPC UA PubSub in industrial automation environments to support the wide variety of traffic types. While the OPC UA PubSub model provides a scalable and decoupled communication mechanism suitable for industrial systems, achieving deterministic message transmission for real-time applications requires a precise definition of traffic specifications and their relevant application configurations. Specifically, the OPC UA specification does not provide explicit mappings between application traffic specifications and the corresponding PubSub configuration parameters. This is especially problematic, as misconfigurations in such an environment can lead to degraded performance, including increased latency or failure to meet critical real-time requirements. This research gap presents a challenge to designing reliable and efficient communication systems in industry. 
In this paper, we address this gap by presenting guidelines for configuring OPC UA PubSub in conjunction with Time-Sensitive Networking (TSN) technologies, thereby enabling real-time guarantees and fault tolerance in industrial communication systems~\cite{zhang_time-sensitive_2024}.
The proposed guidelines assist in mapping industrial automation traffic types to appropriate OPC UA PubSub configurations, ensuring predictable communication and quality of service.
In this context, we first identify a set of industrial automation traffic types commonly used in the automation industry. We then propose a set of guidelines for system and network designers to configure OPC UA PubSub efficiently, accounting for the requirements and characteristics of automation traffic types. We focus on configurations that affect message structure and delivery. Furthermore, this paper examines the publisher’s configuration and assumes that subscribers adhere to it to correctly decode the messages. Subscriber configuration is therefore out of scope of this work and left for future work. Additionally, we compare the configurations proposed in the PubSub specification with those implemented in the open-source OPC UA stack, open62541~\cite{open62541}, and we highlight implementation gaps. 
To validate the proposed guidelines, we present an industrial use case in which the configuration approach is applied, and its impact on network performance is evaluated. This evaluation highlights the practical implications of correct configuration and demonstrates how traffic-aware mapping contributes to predictable communication in real-world scenarios.


\section{Background}
This section describes concepts relevant to understanding the rest of the work, starting with message generation in OPC UA PubSub, continuing with configuration options, and finally delving into UADP messages.
\label{sec:background}
\subsection{OPC UA PubSub Message Generation Overview}
\begin{figure}
    \centering
    \includegraphics[width=1\linewidth]{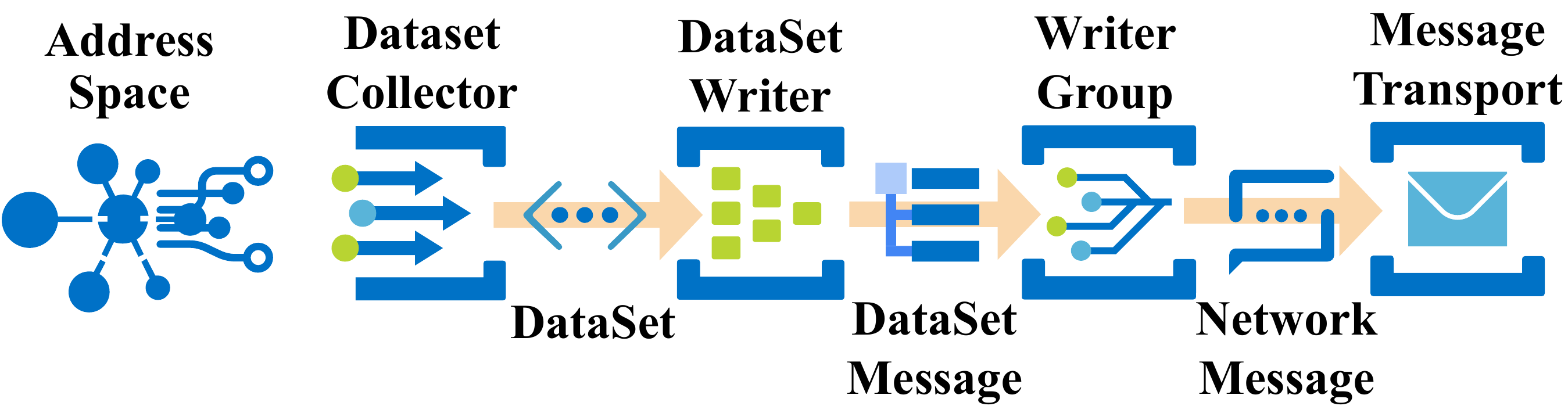}
    \caption{OPC UA PubSub network message generation process and component interactions}
    \label{fig:MessagePublishingProcess}
\end{figure}

To generate PubSub Network Messages (NMs), the DatasetCollector, based on PublishedDataSets (PDSs), selects individual fields such as variables or events from the address space to create Datasets (DSs)~\cite{prinz_configuration_2019, opc_pubsub_2024}, as shown in Fig.~\ref{fig:MessagePublishingProcess}. DS is a list of fields and their metadata. The DataSetWriter (DSW) then uses DSs and message ContentMask to form DatasetMessages (DSMs). Each DSM may contain one or multiple DSs and a header. Next, the WriterGroup (WG) aggregates these DSMs into NMs in every \texttt{PublishingInterval} if they belong to the same group. Finally, the Message Transport layer encapsulates the NM according to the transport protocol specified in the PubSubConnection parameters and transmits the messages over the network. The transport protocol may be broker-based (e.g., MQTT or AMQP) or broker-less, including transmission over User Datagram Protocol (UDP).

\subsection{Dataset Message Types \& Related Configurations}
Each NM in PubSub can be of different types: (i)~Key Frame DSM, (ii)~Delta Frame DSM, (iii)~Event DSM, (iv)~KeepAlive message. There are other message types in PubSub, yet they are not related to data publishing, such as Action Request/Response or Discovery messages. These messages are outside the scope of this work. Following, we present the parameters and configurations that determine message structure and delivery characteristics, as described in the OPC UA PubSub specification - Part 14~\cite{opc_pubsub_2024}.

To differentiate between cyclic and event-based messages, PubSub defines \texttt{CyclicDataset} as a Boolean identifier. This property is part of the PublishedDatasetType component. Derived types such as PublishedDataItemsType and PublishedEventsType represent cyclic and event-based DSs, respectively. In PubSub, cyclic DSMs are typically classified into key frames and delta frames. Key frames include the full DS values and are sent at regular intervals according to the \texttt{PublishingInterval}, regardless of whether there have been changes. 

In contrast, delta frames contain only the changes in values relative to the previous DSM and are not transmitted if no changes occur. To enable key or delta frames, the DatasetMessage Type (DSM Type) parameter should be defined. When using delta frames, it is also necessary to define a \texttt{KeyFrameCount} parameter. When delta frames are enabled, key frames should be configured to be transmitted periodically after a duration equal to \texttt{KeyFrameCount} multiplied by the DS's \texttt{PublishingInterval}. This recurring key frame enables subscribers to synchronise with the publisher. KeepAlive messages, on the other hand, must be transmitted if no key or delta frame has been published within the interval of the \texttt{KeepAliveTime} parameter. \texttt{KeepAliveTime} value should be equal to or greater than the configured \texttt{PublishingInterval}.

The \texttt{DatasetOrdering} parameter defines the order of DSMs in an NM. The first configuration option of \texttt{DatasetOrdering} is Undefined, which specifies the ordering of the DSMs in the NM. The number of DSs in each NM and the number of NMs during \texttt{PublishingInterval} may vary. DSMs are asynchronously grouped based on their readiness at each \texttt{PublishingInterval}. The second one is AscendingWriterID that arranges DSMs by their WriterIDs and packs as many as possible into an NM, constrained by the MaxNetworkMessageSize parameter. In this case, the size and structure of NMs remain consistent as long as the DSM sizes remain unchanged. The third and last option is AscendingWriterIDSingle, which ensures transmission of each DSM in a separate NM.

Furthermore, \texttt{TransportProfileUri} specifies the mapping between the transport protocol and the message encoding used for communication. PubSub supports various transport profiles. Transmission over Ethernet as UA Datagram Protocol (UADP) and JSON over brokers such as AMQP and MQTT. Another parameter that defines the encodings for NMs is \texttt{EncodingMimeType}. The standard defines two available formats, including UADP (Binary) and JSON. 

\subsection{UADP Messages}
In the context of the automation industry and real-time communications, as recommended by the specification~\cite{opc_pubsub_2024}, the UADP message mapping is the preferred choice for encoding and transport. A UADP message consists of the payload containing the DSMs, several possible header sections, including the NM header, GroupHeader (WG-level information), PayloadHeader, one or more DSM headers, and optional headers. In this paper, we mainly focus on UADP message types.


\section{Related Work}
\label{sec:RelatedWorks}
To the best of our knowledge, none of the OPC UA specifications, \cite{opc_uafx_networking_2024}, ~\cite{opc_pubsub_2024}, \cite{OPC_PubSubBaseModel}, or other publications explicitly address traffic type requirements and the translation of these requirements into PubSub configurations. There are a few works that address some of these challenges. For example, in~\cite{opc_uafx_networking_2024}, a recommended hierarchy of network traffic priorities for environments involving Field eXchange stations is introduced. Nonetheless, all traffic originating from OPC UA PubSub is classified under a single category, labeled Green. In practice, however, PubSub traffic can vary significantly in type and criticality. Another mapping outlined in the specifications involves priority-based \ac{QOS}, demonstrating the mechanism of mapping application-layer traffic labels (PriorityLabel) to corresponding \ac{DSCP} and \ac{PCP} values~\cite{opc_pubsub_2024}. This mechanism translates application-layer priorities into associated network-layer priorities via the OPC UA Base Network Model. However, there is no concrete recommendation for application-level traffic and mapping.

Earlier studies on configuring OPC UA PubSub primarily focused on the presentation and organization of PubSub’s standardized components' configuration. In \cite{liu_configuration_2020}, Liu and Bellot implemented an MQTT-based PubSub prototype and an independent configuration tool that utilizes OPC UA view services to retrieve information and configurations from publishers and subscribers. However, the approach does not derive parameters from formal traffic requirements. In parallel, Prinz et al. argued for application-layer configuration submodels for OPC UA PubSub in the Asset Administration Shell~\cite {prinz_configuration_2019}. They focus on structuring the configuration, rather than mapping from application traffic.

Arestova et al. in ~\cite{arestova_service-oriented_2021} attempted to integrate AUTOSAR Adaptive services with OPC UA/TSN to enable heterogeneous, service-oriented communication across domains. They chose to use PubSub for Event messages and other types using the client-server model. They have explicitly mapped each AUTOSAR interface deployment artifact, including identifiers and publishing intervals, to PubSub configuration fields. They also mapped each data type to its corresponding AUTOSAR type. Their mapping is domain-specific and does not derive inputs from heterogeneous traffic specifications.

In summary, while prior works primarily focus on the structural aspects of PubSub's configuration and its integration with other frameworks, they lack systematic guidelines for incorporating traffic specifications in the configuration. This paper addresses such a shortcoming in the literature and practice.


\section{Traffic Specifications to PubSub Configuration Guidelines}
\label{sec:Method}

Our approach to proposing PubSub configurations based on industrial automation traffic specifications begins with a review of existing literature and resources that provide specifications of industrial automation traffic types and PubSub configurations. Our primary basis for analysis and evaluation is the OPC UA Core Specifications, Part 14~\cite{opc_pubsub_2024}, which covers the OPC UA \ac{PubSub} framework. The configuration parameters are also compared to developer guidelines and the open62541 implementation to highlight the inconsistencies, as presented in Table~\ref{tab:pubsub-params}. We then adopt the traffic specification presented in~\cite{automation_TSN_kasra_2025}, which defines 13 distinct traffic types. Required specifications utilized as an input to our guidelines are summarized in Table~\ref{tab:TrafficSpecification}. 

In Table~\ref{tab:TrafficSpecification}, the communication level specification is adopted from the 5G Alliance for Connected Industries and Automation (5G-ACIA)~\cite{5gacia_tsn5g_whitepaper}. Their specification offers a structured categorization of industrial communication traffic types. Moreover, their proposed definitions of communication levels reflect common communication patterns in industrial automation. In this sense, C2C stands for communication between two controllers, C2D represents communication between a controller and field devices, and D2Cmp represents communication from a field device directly to a compute resource, such as an edge server or cloud. In addition, Network traffic type has been removed from the traffic type list in the Table~\ref{tab:TrafficSpecification}, since the source of these messages is not an OPC UA publisher, and they originate from network devices. Voice and Video types share the same characteristics and are grouped under one category. 

To propose comprehensive configurations, all components involved in the message generation procedure must be systematically assessed. The configurations are derived from Server, WG, PubSubConnection, PDSs, and DSW components. To recommend a valid PubSub configuration, particular attention is given to message generations and properties that influence the structure and behavior of NetworkMessages (NMs), which is the publisher's role and configuration. Furthermore, subscribers' configurations must align with the corresponding publisher's configurations to decode the received messages. Therefore, evaluating subscribers' configurations is left for future work and not in the scope of this paper. Next, we propose each selected PubSub parameter relevant to the message generation process, based on the traffic specification.

\begin{table*}[!h]
\centering
\caption{Identified PubSub configuration parameters, including their base components, standard definition status, and stack-level status}
\label{tab:pubsub-params}
\small
\setlength{\tabcolsep}{4pt}
\renewcommand{\arraystretch}{1.2}
\begin{tabular}{|>{\raggedright\arraybackslash}p{3.5cm}|>{\raggedright\arraybackslash}p{2.4cm}|>{\raggedright\arraybackslash}p{2.8cm}|>{\raggedright\arraybackslash}p{2.2cm}|>{\raggedright\arraybackslash}p{4cm}|}
\hline
\textbf{Parameter} & \textbf{Base Component} & \textbf{Standard Definition} & \textbf{Implementation} & \textbf{Stack Parameter Name}\\
\hline
DSM Type                  & DSMHeader         & Yes & Stack   &                          \\ \hline
KeyFrameCount             & DSW               & Yes & Stack   &                          \\ \hline
CyclicDataset             & PDS               & Yes & No      & PublishedDataSetType     \\ \hline 
KeepAliveTime             & WG                & Yes & Stack   &                          \\ \hline
PublishingInterval        & WG                & Yes & Stack   &                          \\ \hline
EncodingMimeType          & WG                & Yes & Stack   &                          \\ \hline
DatasetOrdering           & WG                & Yes & No      & MaxEncapsulatedDSMCount  MaxNetworkMessageSize\\ \hline
TransportProfileUri       & PubSubConnection  & Yes & Stack   &                          \\ \hline
EnableDeltaFrames         & Server            & No  & Stack   &                          \\ \hline

\end{tabular}
\end{table*}

\begin{table*}[!h]
\centering
\caption{Industrial automation traffic types and their communication characteristics~\cite{automation_TSN_kasra_2025}}
\renewcommand{\arraystretch}{1.2}
\small
\begin{tabular}{|c|>{\raggedright\arraybackslash}p{2.4cm}|>{\raggedright\arraybackslash}p{1cm}|>{\raggedright\arraybackslash}p{1.3cm}|>{\raggedright\arraybackslash}p{2.1cm}|>{\raggedright\arraybackslash}p{2.6cm}|>{\raggedright\arraybackslash}p{3.6cm}|}
\hline
\textbf{ID} & \textbf{Traffic ID} & \textbf{Periodic} & \textbf{Criticality} & \textbf{Loss Tolerance} & \textbf{Length consistency} & \textbf{Communication Level} {\scriptsize\cite{5gacia_tsn5g_whitepaper}} \\
\hline
1 & Control-Iso        & Yes & High   & No  & Fixed    & C2C, C2D        \\
\hline        
2 & Control-Sync       & Yes & High   & No  & Fixed    & C2C, C2D        \\
\hline    
3 & Control-Async      & Yes & High   & Yes & Fixed    & C2C, C2D        \\
\hline
4 & Event              & No  & High   & Yes & Variable & C2C, C2D, D2Cmp \\
\hline
5 & Voice/Video        & Yes & Low    & Yes & Variable & D2Cmp           \\
\hline
6 & Command-Cycle      & Yes & Medium & Yes & Variable & C2D, D2Cmp      \\
\hline
7 & Command-Acycle     & No  & Medium & Yes & Variable & C2D, D2Cmp      \\
\hline
8 & Config             & No & Medium  & Yes & Variable & C2C, C2D, D2Cmp \\
\hline
9 & Diagnostic-Cycle   & Yes & Medium & Yes & Variable & C2C, C2D, D2Cmp \\
\hline
10 & Diagnostic-Acycle & No & Medium  & Yes & Variable & C2C, C2D, D2Cmp \\
\hline
11 & Best-Effort       & No & Low     & Yes & Variable & D2Cmp           \\
\hline
\end{tabular}
\label{tab:TrafficSpecification}
\end{table*}

\textbf{Cyclic and Event-based Messages:} For periodic DSs other than setting the \texttt{CyclicDataset}, the DSM Type should be set to key frame. Additionally, \texttt{PublishingInterval} should be set relative to their cycle time. For Event-based DSs, in addition to setting \texttt{CyclicDataset} to False, the DSM Type should be set to Event.

\textbf{Delta Frame Messages:} It is recommended to use delta frame messages when value updates are cyclic in nature and should be avoided for event-based data~\cite{opc_pubsub_2024}. The application consuming delta frames should be designed to handle data loss, where occasional packet loss does not impact its functionality, as packet loss in delta frames may cause inconsistency, making them unsuitable for control messages. For lost messages, the standard only mentions the MessageReceiveTimeout mechanism as part of the subscriber's configuration, which leaves handling of missed frames to the implementation. The problem arises if the subscriber misses both a key frame and one or more delta frames.

Eventually, the subscriber loses track of the messages, since further messages might be delta frames (based on the \texttt{KeyFrameCount} configuration). In this condition, the system loses integrity and determinism until it receives the next key frame. While employing delta frames, \texttt{KeyFrameCount} should be defined based on the use case, with a value greater than 1. The default value of 1 represents that the key frames should be transmitted at each \texttt{PublishingInterval}.

Moreover, delta frames are not suited for media streams such as voice or video. The loss of a key frame or a single packet in transmitting byte streams can render subsequent delta frames undecodable due to their dependency on preceding reference data. The best fit for voice and video messages that transmit surveillance data in PubSub is to utilize chunk messages, which facilitate sequential reconstruction and can handle streams. In contrast, the transmission of images from industrial cameras presents a different scenario: these payloads are relatively constrained in size and are typically designed for downstream image processing rather than continuous streaming. The traffic type for these messages, given their smaller size and higher criticality, would be either Control-Iso, Control-Sync, or Control-Async.

Another situation in which the potential cost of delta frames may exceed the gains is when we expect the cyclic changes to be continuous. In such cases, the potential computational overhead of field comparisons can outweigh the benefits of reducing bandwidth usage, since the source of these messages is typically a resource-constrained embedded system. In this case, the potentially utilized bandwidth may increase compared to using key frames. This is due to the nature of delta frames, which carry index information for each field. The impact on bandwidth is tied to the transmission cycle of these continuously changing data, which determines how much indexed data needs to be transmitted.

\textbf{KeepAlive Message:} It is recommended to use KeepAlive messages for cyclic data that employs delta frames. Generally, keepAlive messages are crucial for informing subscribers when data stops arriving for any reason, e.g., due to no updates or a failure.  When there is no update, the subscriber can rely on the key frames transmitted according to the configured \texttt{KeyFrameCount}. However, if the delta frames are missed in between, the subscriber has no reference to detect this without employing keepAlive messages.  It helps reduce bandwidth usage for data sources that change infrequently by avoiding unnecessary transmissions (cyclic data transmitted on change) and speeding up error detection. Instead of sending more frequent key frames, the subscriber relies on \texttt{KeepAliveTime} to ensure the previous value remains valid. For event-driven messages, employing KeepAlive messages is also critical, as it allows subscribers to detect and report communication interruptions. In the case of publisher failure, the subscriber can rely on keepAlive messages for any further decision. Finally, it is not recommended to use KeepAlive messages for cyclic DSs that do not use delta frames. Since, by the time the cyclic messages stop arriving, the issue can typically be attributed to a problem at the data source or a network failure.

\textbf{EncodingMimeType:} The PubSub specifications recommend using UADP binary encoding for scenarios that require deterministic transmission. By applying normative UADP layouts defined in the specifications, it is possible to construct fixed-sized messages with minimal header overhead that support the deterministic requirements. In this sense, UADP is suitable for deterministic and control communications, such as controller-to-controller and controller-to-field device communication, as it has less encoding overhead than JSON. JSON should be used for communication with enterprise-level and cloud systems or environments that do not support PubSub's UADP encoding.

\textbf{DatasetOrdering:} This parameter depends on the criticality, length consistency, and the communication level of the traffic. The first option, named Undefined, is suitable for bulk D2Camp transmissions without strict timing guarantees, as it reduces network bandwidth usage and computational overhead. However, this option is not suitable for traffic types with deterministic transmission requirements, as there is no consistency in the NM size. AscendingWriterID or AscendingWriterIDSingle is appropriate when the publisher transmits multiple cyclic DSMs. If there is only one dataset originating from the publisher, it does not matter which option will be chosen.

AscendingWriterID helps minimize bandwidth and computational overhead by aggregating DSMs according to \texttt{PublishingIntervals}. However, if this message were sent to many subscribers who are only interested in a subset of the DSMs, it might increase the bandwidth. This happens when at least one subscriber is interested in all DSMs of the same Publisher. For deterministic fixed-size NMs, this option is recommended. Although DSs tend to group together in real-world scenarios, the AscendingWriterIDSingle is recommended for any scenario in which a single fixed-size DSM should be published or when the \texttt{PublishingInterval} of the DSMs varies. This is important to ensure deterministic transmission for time-critical data. In options Undefined and AscendingWriterID, message ordering may be influenced by the MaxNetworkMessageSize parameter. Finally, for critical event-based messages, to reduce the waiting time until the next Publishing interval of the group, AscendingWriterIDSingle is the recommended choice.

\textbf{TransportProfileUri:} For time-sensitive, broker-less industrial communication, OPC UA defines the PubSub UDP UADP transport profile, which enables low-latency and deterministic data exchange using UDP multicast or unicast. In this paper, we adopt this recommended transport profile to transmit UADP messages over the UDP protocol~\cite{opc_pubsub_2024}. The underlying network stack would be TSN in this case. In scenarios where there is no real-time constraint or the receiver does not support PubSub, JSON can be adopted.

Inconsistencies among the standard definitions of the presented parameters and the open62541 implementation start with configuring delta frames. Open62541 includes an additional parameter, \texttt{EnableDeltaFrames}, in the global server component that follows the same logic as defining the DSM Type as Delta Frame. Regarding the DSM Type, the specification does not specify to which component it belongs. They mention DSM Type as part of the DatasetMessage header structure. On the other hand, open62541 defines the DSM Type parameter and follows the same logic as explained in the standard. 

Although the PubSub specification defines the \texttt{DatasetOrdering} parameter, it is not yet implemented in open62541. Instead, the open62541 represents other parameters, such as MaxEncapsulatedDSMCount and MaxNetworkMessageSize, which specify the maximum number of DSMs in an NM and the maximum NM size. With these configurations, one can simulate the AscendingWriterIDSingle. However, the encapsulation of DSs in an NM will follow the natural ordering of the DSs defined in the code. To simulate cyclic and event DSs, the PublishedDataSetType and \texttt{CyclicDataset} parameters are described in the standard. However, open62541 only uses PublishedDataSetType. Other parameters have the same name and implementation logic.

Table \ref{tab:MessageTypeConfiguration} summarizes our guidelines for corresponding  DSM Types and their related configurations for each industrial automation traffic type. Dependent in this table means that if a delta frame is employed, the value is Yes, and \texttt{KeyFrameCount} is to be greater than 1; otherwise, the value should be No, and \texttt{KeyFrameCount} should be equal to 1. Regarding the \texttt{DatasetOrdering} parameter, Undefined is considered as 1, AscendingWriterID is considered as 2, and AscendingWriterIDSingle is considered as 3. 

\begin{table*}[!ht]
\centering
\caption{The Guideline of Industrial Automation Traffic Types to PubSub Configurations}
\renewcommand{\arraystretch}{1.2}
\small
\begin{tabular}{|>{\raggedright\arraybackslash}p{1cm}|>{\raggedright\arraybackslash}p{3.2cm}|>{\raggedright\arraybackslash}p{1.3cm}|>{\raggedright\arraybackslash}p{0.9cm}|>{\raggedright\arraybackslash}p{1.5cm}|>{\raggedright\arraybackslash}p{1.3cm}|>{\raggedright\arraybackslash}p{1.7cm}|>{\raggedright\arraybackslash}p{1.6cm}|}
\hline
\textbf{ID} & \textbf{DSM Type} & \textbf{KeyFrame Count} & \textbf{Cyclic Dataset} & \textbf{EnableDelta Frames} & \textbf{KeepAlive Time} & \textbf{Encoding \& Transport} & \textbf{Dataset Ordering} \\
\hline
1, 2 , 3 & Key Frame                     & 1           & Yes    & No        & No           & UADP         & 2 or 3  \\ \hline
4        & Event                         & 1           & No     & No        & Yes          & UADP         & 3 or 1  \\ \hline
5        & Chunk Message                 & 1           & Yes    & No        & No           & UADP         & 1       \\ \hline
6        & Key Frame/ Delta Frame        & 1 or $>1$   & Yes    & Dependent & Dependent    & UADP/JSON    & 1       \\ \hline
7        & Event                         & 1           & No     & No        & Yes          & UADP/JSON    & 1       \\ \hline
8        & Event                         & 1           & No     & No        & Yes          & UADP/JSON    & 1       \\ \hline
9        & Key Frame/ Delta Frame        & 1 or $>1$   & Yes    & Dependent & Dependent    & UADP/JSON    & 1       \\ \hline
10       & Event                         & 1           & No     & No        & Yes          & UADP/JSON    & 1       \\ \hline
11       & Event                         & 1           & No     & No        & Yes          & UADP/JSON    & 1       \\ \hline
\end{tabular}
\label{tab:MessageTypeConfiguration}
\end{table*}

\section{Evaluation}
\label{sec:Evaluation}
This section presents the qualitative evaluation of the proposed guidelines using an industrial automation use case. We first describe the use case considered and its relevance, and then we present the qualitative evaluation.

\subsection{Use Case Scenario}
We consider an industrial communication use case in which two production lines interact with higher-layer systems, such as edge, SCADA, and cloud services. The setup reflects a typical Industry 4.0 architecture similar to the one presented in~\cite{selim_Resilient_2025}, where time-sensitive machine-to-machine communication coexists with higher-level monitoring and cloud communication. All devices communicate via broker-less OPC UA PubSub over UDP (unicast or multicast), and switches are TSN-enabled.

\begin{figure}
    \centering
    \includegraphics[width=1\linewidth]{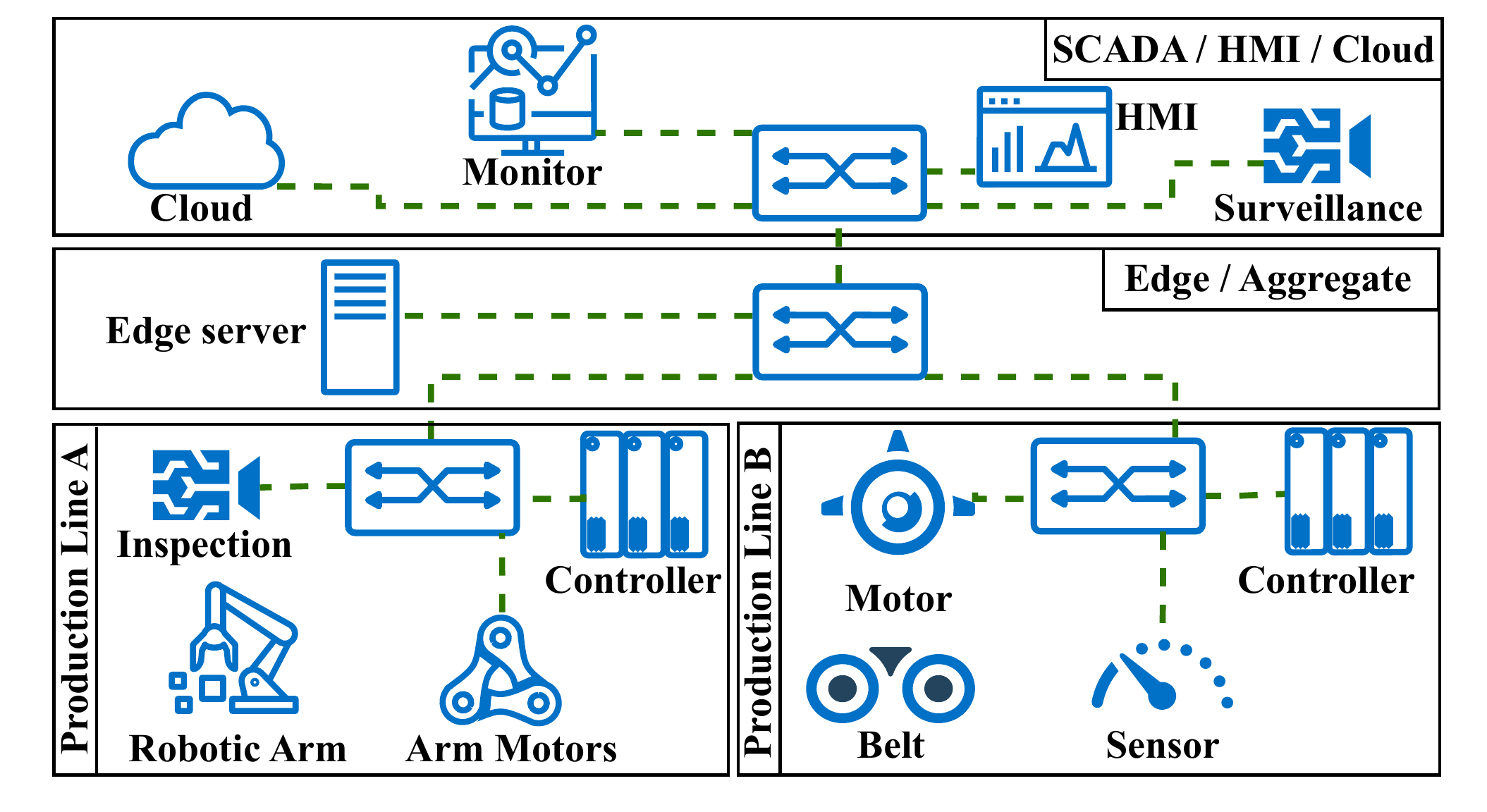}
    \caption{Industrial automation use case illustrating communication between field devices, edge node, and cloud services in four zones}
    \label{fig:usecase}
\end{figure}

Considering the use case depicted in Fig.~\ref{fig:usecase} and the traffic flows presented in Table~\ref{tab:TrafficFlow}, line A consists of a robotic arm equipped with a motor. The robotic arm's actuation and next movement are dependent on the output of an industrial camera. The camera sends Control-Sync data to the edge server (Flow 6), where computationally demanding tasks are performed. This design offloads high-computational tasks that the controller A cannot manage to the edge. After processing the data, the edge server publishes refined commands to controller A (Flow 8) and reports additional data to the cloud (Flow 9) for further analysis. During this communication, telemetry or status data published by the edge server can be processed by a monitoring agent for diagnostics. In addition, the arm's motor transmits Iso-control messages to controller A, as shown in Flow 7.

Line B contains a conveyor belt with a motor and a sensor, managed by controller B. In contrast to Line A, the conveyor’s movement depends on the information provided by its sensor B, the belt's motor output, and controller A's output. In this case, the processing requirement is less intensive, and there is no communication to the edge. Controller A sends Control-Sync messages to Controller B (Flow 1) to synchronize the conveyor belt and the robotic arm. These messages are transmitted at very short intervals, ranging from hundreds of microseconds to a few milliseconds. Sensor B publishes Control-Async messages containing the measured distance to objects detected on the conveyor belt. Upon detecting an object, it publishes this data, and Controller B, as one of the subscribers, consumes it for processing. 

In addition, all motors, sensors, and controllers in the system publish event or cyclic messages that include status updates, warnings, and error messages, as shown in Flow 4 and Flow 5. These messages are consumed by various nodes, including the monitoring and HMI node, which relies on them for diagnostics and system health tracking.

\begin{table*}[!ht]
\centering
\caption{Traffic Flow Among Nodes in Industrial Automation Use Case}
\renewcommand{\arraystretch}{1.2}
\small
\begin{tabular}{|>{\raggedright\arraybackslash}p{1.2cm}|>{\raggedright\arraybackslash}p{3.7cm}|>{\raggedright\arraybackslash}p{1.8cm}|>{\raggedright\arraybackslash}p{3cm}|>{\raggedright\arraybackslash}p{4.1cm}|}  
\hline
\textbf{Flow ID} & \textbf{Publisher} & \textbf{Subscriber} & \textbf{Traffic Type}  & \textbf{Purpose}  \\
\hline
Flow 1 & Controller A                 & Controller B & Control-Sync                 & Arm and belt synchronization \\ \hline
Flow 2 & Sensor B                     & Controller B & Control-Async                & Movement detection on belt \\ \hline
Flow 3 & Controller B                 & HMI          & Command-Cyclic               & Movement detection on belt \\ \hline
Flow 4 & Motors, Sensors, Controllers & HMI Monitor  & Event                        & Diagnostics and updates \\ \hline
Flow 5 & Arm Motor                    & Monitor      & Command-Cyclic  & Diagnostics \\ \hline
Flow 6 & Camera A                     & Edge Server  & Control-Sync                 & Arm's movement computations \\ \hline
Flow 7 & Arm Motor                    & Controller A & Control-Iso                  & Sensory data updates \\ \hline
Flow 8 & Edge Server                  & Controller A & Control-Sync                 & Arm's movement results \\ \hline
Flow 9 & Edge Server                  & Cloud        & Best-Effort                  & Historian and Analysis \\ \hline
\end{tabular}
\label{tab:TrafficFlow}
\end{table*}

As stated in the PubSub specification, the header layouts are designed to be flexible and support a wide range of use cases, but this flexibility also increases implementation and verification complexities~\cite{opc_pubsub_2024}. By adopting the normative header proposed in Annex A of~\cite{opc_pubsub_2024}, we align with predefined layouts specified by the specification to reduce complexity. Depending on the length consistency of each traffic type, either a Periodic Fixed Header Layout or a Dynamic Header Layout is proposed. When security mechanisms are required, the same layouts with signing and encryption should be applied. In scenarios involving a combination of different DSMs published by the same publisher, additional considerations, such as DSM grouping and their impact on the NM, must be taken into account and are beyond the scope of this paper. In the fixed layout, each NM within a given \texttt{PublishingInterval} should contain the same number of DSMs, and each DSM should include the same number of DSs. This constraint enables several optimizations and reductions in header size.


\subsection{Evaluation Based on the Use Case}

\subsubsection{Dataset Message Type \& EnableDeltaFrames}
In this section, we cover both the DSM Type and the \texttt{EnableDeltaFrames} parameter breakdowns for OPC UA PubSub. One of the critical decisions in PubSub configuration is selecting between key frames and delta frames as the DSM Type. At the same time, the distinction between key frames and event messages is relatively straightforward and discussed in Section \ref{sec:Discussion-Cyclic}. Consider Flow 1; if the DSM Type is set to delta frame, only the changes in the DS are transmitted. The first challenge to arise is frame loss. If a packet is lost, Controller B may lack a valid reference (i.e., the last key frame) to interpret the incoming delta frame. Update time of the DSs increases due to a loss of delta or a key frame. Apart from this, since the specifications leave handling such scenarios to the end user, it makes it vendor-specific, which is against the heterogeneous characteristic of the OPC UA PubSub. Even if the system implements mechanisms such as sequence number tracking, it may only be sufficient to detect the error and not to reconstruct the lost packet and its DS, which makes defining \texttt{KeyFrameCount} crucial here. While packet loss can occur for both key frames and delta frames, the key advantage of key frames is that any subsequent key frame received by Controller B serves as a complete reference, which resynchronizes the DS and resolves any ambiguity caused by the lost packet.

The second challenge, however, concerns the high update frequency. Given the rapid transmission rate and the dynamic nature of control parameters, the DS fields are constantly changing. In such cases, delta frames are inefficient. Each delta frame must include not only the updated fields but also a 2-byte field index for every changed field, which introduces additional overhead. In contrast, key frames, despite transmitting the whole DS, are more bandwidth-efficient under these conditions due to their simpler structure and reduced parsing complexity.

In another scenario, Flow 2 is transmitted as key frames. In addition to Controller B, the HMI requires access to sensor data. However, they do not require updates at the same frequency as Controller B requires them. To optimize system performance and reduce unnecessary network load, Controller B serves as an intermediary publisher, forwarding relevant data to the HMI as Flow 3. The HMI is interested in state changes rather than continuous data streams. Therefore, Controller B transmits Control-Async delta frames rather than key frames. If key frames were used in this HMI scenario, it would result in higher bandwidth utilization.

Considering Flow 4, suppose these event-driven data are incorrectly configured to publish as key frames. In that case, the OPC UA publisher will transmit them cyclically, regardless of whether any state change has occurred. This behavior deviates from the intended event-based communication model. Such a misconfiguration can lead to excessive network bandwidth consumption. Moreover, it causes a risk of propagating stale or misleading information.

\subsubsection{KeyFrameCount}
Flow 5 updates are transmitted using delta frames to minimize bandwidth traversing the entire network. However, with a misconfigured \texttt{KeyFrameCount} set to a high value, the monitoring node is left in an ambiguous state. It cannot distinguish whether (i) the reported value has not changed, or (ii) there is a failure in the publisher. This uncertainty compromises the monitoring node's ability to assess the arm motor's health, even when the communication links among them remain active. This problem arises when the delta frame is utilized, such as Command-Cycle or Diagnostic-Cycle traffic. However, the specification defines the MessageReceiveTimeout parameter as a subscriber configuration, which should be properly configured in relation to \texttt{PublishingInterval}, \texttt{KeepAliveTime}, and \texttt{KeyFrameCount} to prevent this situation. In contrast, in scenarios where no delta frames are employed, consider Flow 6 transmitting key frames; the definition of the \texttt{KeyFrameCount} parameter becomes irrelevant, which has no impact on the system.

\subsubsection{Cyclic Dataset}
\label{sec:Discussion-Cyclic}
\texttt{CyclicDataSet} mode ensures that the publisher transmits NMs on every tick of the \texttt{PublishingInterval}, regardless of data changes. Having Flow 7 transmitting key frames, if the \texttt{CyclicDataSet} parameter is incorrectly set to false for cyclic traffic, the publisher will treat the messages as event-driven rather than periodic. In this mode, the motor's publisher does not enforce a fixed schedule or interval; it transmits only when a data change is detected. If the data source (e.g., torque) updates exactly at the expected transmission interval, the subscriber may perceive no difference since the stream appears periodic. Even though the DSM Type is an event, the payload structure remains the same as a key frame. However, the data source may deviate from the expected rate, leading to a loss of deterministic isochronous traffic transmission. This may lead to missed messages, timeouts, or degradation of data quality at the controller (subscriber).

Furthermore, the deviation disrupts network scheduling and introduces jitter and delay in time-triggered networks such as TSN. In event mode, DSW only inserts the DS into the next NM based on DS's \texttt{PublishingInterval} once a change is observed. If the change arrives a few microseconds after the WriterGroup’s scheduled tick, the update slips to the next cycle, causing periodicity drift.

\subsubsection{KeepAliveTime}
If the monitoring node and motor B are using Flow 4 to transmit threshold-crossing event frames, the monitor node still needs a way to verify that the motor node is alive. To address this, keepAlive messages can be incorporated into the PubSub mechanism. Without keepAlive messages, custom heartbeat handling would be required, where the controller periodically informs the monitor node of the motor’s availability. When a publisher advertises a keepAlive message, all its subscribers are notified that the node is active, even when no events occur. Besides event-driven messages, this approach is also beneficial in scenarios where delta frames are transmitted infrequently between nodes. On the other hand, enabling keepAlive messages in Flow 8 transmitting key frames provides no additional benefit other than consuming redundant bandwidth. The moment the device stops receiving key frames, it concludes that a failure has occurred either in the network or on the publisher. 

\subsubsection{EncodingMimeType \& TransportProfileUri}
Flow 9 is conducted using JSON encoding, primarily because most cloud platforms are compatible with typical RESTful APIs and lack native PubSub support. Employing the UADP encoding and transport profile for cloud communication introduces more complexity to the system. This includes challenges in serialization and protocol translation across heterogeneous environments. Moreover, not using OPC UA's models for such communication decouples the systems from the OPC UA information model and limits the semantic interoperability and structured data representation in OPC UA.

Moreover, using JSON instead of UADP for communication, when deterministic transmission is required, results in greater delays in both encoding and decoding compared to the direct binary mapping of fields to memory in UADP. Furthermore, JSON messages are verbose and include field names, thereby increasing bandwidth consumption.

\subsubsection{DatasetOrdering}
To support our argument, consider Flow 3. Configuring the \texttt{DatasetOrdering} to AscendingWriterIDSingle, which should be set to Undefined, forces the controllers to publish an NM at each \texttt{PublishingInterval} for each DS. However, in fact, the controllers have numerous DSs with varying \texttt{PublishingInterval}s to publish to these nodes, including motor status, sensor data, and the controller's own status. Thus, transmitting an NM for each DS incurs considerable overhead for devices and the network. Therefore, although the messages may have different \texttt{PublishingInterval}s, selecting the Undefined option increases the likelihood that DSs from multiple WGs will be combined into the same NM, thereby reducing the previously discussed overhead. In scenarios where \texttt{PublishingInterval}s are equal, you can indeed publish DSs in the controller simultaneously and group them under the same WG, thus they will be published as a unit. Best-Effort transmissions to the cloud also exhibit the same characteristics, such as historian updates. 

On the other hand, if you choose the Undefined option for Flow 7, you are essentially breaking determinism, which is one of the requirements of real-time communications by imposing variable-length NMs. Time-sensitive networks require a fixed message size to schedule the network specifically at the field level and control communication. 

It should be noted that the overhead we discuss includes the devices' overhead (decoding and encoding) and network utilization overhead. These overheads increase if the security headers are employed.

\section{Conclusion and Future Works}
\label{sec:conclusion}
In this paper, we introduced systematic guidelines for mapping industrial automation traffic types to OPC UA PubSub configurations based on their timing and quality-of-service specifications. These guidelines were derived from an analysis of traffic characteristics and their influence on communication parameters. We validated the approach through an industrial use case, demonstrating how misconfiguration can lead to significant performance degradation, and showing that applying the proposed guidelines improves network predictability and stability.
Future work aims to conduct experiments to assess performance metrics using the proposed guidelines in industrial settings. Moreover, we aim to investigate the interactions between OPC UA configuration parameters and different TSN traffic shaping mechanisms to further enhance real-time guarantees and fault tolerance in Industry 4.0 systems.

\section*{Acknowledgments}
This work is supported by the Swedish Knowledge Foundation (KKS) via the SEINE and MARC projects and by the Swedish Governmental Agency for Innovation Systems (VINNOVA) via the FLEXATION and iSecure projects. The authors thank the industrial partners, in particular, ABB, Westermo, and Arcticus Systems, for providing their valuable input to this work.


\bibliographystyle{IEEEtran}
\bibliography{references}

\end{document}